\documentclass[a4paper,11pt]{article}
\pdfoutput=1 

\usepackage{jheppub} 

\numberwithin{equation}{section}

\gdef\ffrac#1#2{\textstyle\frac{#1}{#2}\displaystyle}
\gdef\be{\begin{equation}}
\gdef\ee{\end{equation}}
\gdef\e{\epsilon}
\gdef\a{\alpha}
\gdef\p{\partial}

\title{\boldmath $T\overline T$ deformations of non-Lorentz invariant field theories}

\author[a,b]{John Cardy}

\affiliation[a]{Department of Physics, University of California, Berkeley CA 94720, USA}
\affiliation[b]{All Souls College, Oxford OX1 4AL, UK}

\emailAdd{cardy@berkeley.edu}

\abstract
{We point out that the arguments of Zamolodchikov and others on the $T\overline T$ and similar deformations of two-dimensional field theories may be extended to the more general non-Lorentz invariant case, for example non-relativistic and Lifshitz-type theories. We derive results for the finite-size spectrum and $S$-matrix of the deformed theories.  
}

\begin{document} 
\maketitle

\section{Introduction}\label{sec:intro}

In  2004 Zamolodchikov \cite{Zam} obtained analytic results for the expectation value of the operator 
${\cal O}\equiv T_{zz}T_{\bar z\bar z}-T^2_{z\bar z}$ in a two-dimensional relativistic quantum field theory, where the $T_{\mu\nu}$ denote the components of the (euclidean) energy-momentum, or stress, tensor in complex coordinates. In particular he showed that it is well-defined by point-splitting, and that its vacuum expectation value is proportional to $-\langle T^\mu_\mu\rangle^2$. 

Furthermore he showed that, when the field theory is compactified on a spatial circle of circumference $R$, the expectation value $\langle n|{\cal O}|n\rangle$ in any eigenstate $|n\rangle$ of energy and momentum is  related to the expectation values  $\langle n|T^\mu_\nu|n\rangle$. This leads to a simple differential equation for how the energy spectrum at finite $R$ evolves on deforming the action by a term $\propto {\cal O}$. 

This has become known as the $T\overline T$ deformation of the original two-dimensional local field theory, and in recent years has become the object of some attention, one reason being that it turns out to be an example of a local field theory perturbed by irrelevant (non-renormalizable) operators which nevertheless has a sensible UV completion which is not, however, itself a local QFT. It has been argued that this deformation corresponds to a modification of the $S$-matrix by CCD factors \cite{Dub00,Dub0,Smi}, and corresponds \cite{Dub,Dub2} to a dressing of the theory by Jackiw-Teitelboim \cite{Jac,Tei} gravity. The deformation of a free boson theory was shown \cite{Cas} to correspond to the spectrum of the Nambu-Goto string, and like that theory, for one sign of the deformation it has a Hagedorn transition at finite temperature. The connection between the deformed $S$-matrix and the finite-size spectrum was verified in a number of integrable models \cite{Cav}. It was also argued \cite{Ver} that the opposite sign of the coupling is equivalent  to going into the bulk in the context of the AdS-CFT correspondence. The effect on scattering at finite energy density was discussed in \cite{Car,Ver,Ber}. 
Since these papers appeared the subject has received considerable attention: 
\cite{Gui,Giv,Shy,Gir,Kra,Tay,Aha,Bon,Dat,Aha2} are relevant to the present work.

In a previous paper \cite{Car2} it was argued that the solvability of this deformation (which should be more correctly termed
$\det T$, since it reduces to $T\overline T$ only in the case of a CFT) may be explained in terms of coupling the theory to a random metric, whose action is itself topological. In that paper, Zamolodchikov's \cite{Zam} original argument (which he gave in complex coordinates) was re-expressed in Cartesian coordinates. In this framework it is straightforward to see that his arguments, and those of \cite{Car2}, generalize to non-Lorentzian invariant theories and more general deformations thereof.
It is the purpose of this short note to document this.

\section{\boldmath $\det T$ deformation}

As usual, we consider a sequence of theories ${\cal T}^{(t)}$ on flat 2d space labelled by a real parameter $t$. ${\cal T}^{(0)}$ is a local quantum field theory, although not necessarily Lorentz invariant (\em i.e.\em\ not rotationally invariant in Euclidean signature). We use Cartesian coordinates $(x_0,x_1)$ where $x_0$ is (Euclidean) time and $x_1$ is space, and the Euclidean metric is $\delta_{ij}$ so we do not need to distinguish upper and lower indices. 

We assume translational invariance in both coordinates, with generators $H$ and $P$ respectively when the theory is quantized on a constant $x_0$ interval. We assume that ${\cal T}^{(t)}$ possess a local energy-momentum (stress) tensor with components $T^{(t)}_{ij}$, which  may be viewed as the response to a strain field $\a_{i,j}$ induced by a diffeomorphism $x_i\to x_i+\a_i(x)$. In particular 
\be
H=\int T^{(t)}_{00}dx_1\,,\quad P=i\int  T^{(t)}_{10}dx_1\,,
\ee
where $T^{(t)}_{00}$ and $T^{(t)}_{10}$ are the energy and momentum densities respectively. The stress tensor is conserved: $\p_jT_{ij}=0$
\be
\p_0T_{00}+\p_1T_{01}=\p_0T_{10}+\p_1T_{11}=0\,,
\ee
where $iT^{(t)}_{01}$ and $T^{(t)}_{11}$ are the energy and momentum fluxes respectively. However we do not assume that $T^{(t)}_{10}=T^{(t)}_{01}$, which would correspond to invariance under Euclidean rotations (Lorentz boosts). Note that such theories (for example non-relativistic fluids) may have additional conserved currents which may be incorporated into an enlarged stress-energy-momentum-mass tensor, but we shall not consider this here. Our point of view is that ${\cal T}^{(0)}$ is an emergent theory describing, for example, the long-wavelength low energy behavior of a condensed matter system with no further symmetries. In addition we assume there is no dissipation so the energy flux is the same as the heat flux. 

The deformation is defined incrementally from ${\cal T}^{(t)}$ to ${\cal T}^{(t+\delta t)}$ by formally adding a term
\be\label{1.3}
-2\delta t\int e_{ik}e_{jl}T^{(t)}_{ij}(x)T^{(t)}_{kl}(x)d^2x
\ee
to the action, or, less formally, by inserting it into correlation functions. Here $e$ is some $2\times 2$ matrix which may itself depend on $t$. The two tensors $e_{ik}$ and $e_{jl}$ should be identical however, because of the symmetry under $(ij)\leftrightarrow(kl)$. Lorentz invariance would demand that $e_{ik}$ is proportional to the Levi-Civita tensor $\e_{ij}$, in which case the deformation is $\propto \det T^{(t)}$ and we recover the standard (so-called) $T\overline T$ deformation as discussed extensively in the literature. 

However, following the version of Zamolodchikov's argument \cite{Zam} given in \cite{Car2}, we may consider the variation of the point-split version of (\ref{1.3}) (we drop the $(t)$ superscripts for clarity)
\be\label{1.4}
\p_{y_m}e_{ik}e_{jl}T_{ij}(x)T_{kl}(x+y)\,.
\ee
To proceed we require that the following identity holds:
\be
e_{jl}\p_{y_m}=e_{ml}\p_{y_j}+e_{jm}\p_{y_l}\,.
\ee
By enumerating the different cases, it is straightforward to see that $e$ must be proportional to the Levi-Civita symbol $\e$. The constant of proportionality may be absorbed into $\delta t$ and henceforth we use $\e$ rather than $e$. 

Thus we may write (\ref{1.4}) as
\begin{eqnarray}
&=&\p_{y_j}\e_{ik}\e_{ml}T_{ij}(x)T_{kl}(x+y)+\p_{y_l}\e_{ik}\e_{jm}T_{ij}(x)T_{kl}(x+y)\\
&=&\e_{ik}\e_{ml}T_{ij}(x)\p_{x_j}T_{kl}(x+y)=\p_{x_j}[\e_{ik}\e_{ml}T_{ij}(x)T_{kl}(x+y)]\,,
\end{eqnarray}
where we have used $\p_lT_{kl}=0$ and $\p_jT_{ij}=0$.

Thus, in any translationally invariant state,
\be\label{1.7}
\e_{ik}\e_{jl}\langle n| T_{ij}(x)T_{kl}(x)|n\rangle=\e_{ik}\e_{jl}\langle n| T_{ij}(x)T_{kl}(x+y)|n\rangle\,,
\ee
for all $y$.

Zamolodchikov's argument \cite{Zam} proceeds by considering the theory quantized on the periodic interval $0\leq x_1\leq R$, and inserting a complete set of eigenstates $\sum_m|m\rangle\langle m|$
of $H$ and $P$ on the right hand side 
of (\ref{1.7}). The fact that the result should be independent of $y$ then implies that $(E_m,P_m)=(E_n,P_n)$.
Within each eigenspace we may then choose a basis in which $T_{00}\propto H$ and $T_{10}\propto P$ are diagonal, and thus without loss of generality take $|m\rangle=|n\rangle$. (\ref{1.7}) then becomes
\be
\langle n|T_{00}T_{11}-T_{01}T_{10}|n\rangle
=\langle n|T_{00}|n\rangle\langle n|T_{11}|n\rangle-
\langle n|T_{01}|n\rangle\langle n|T_{10}|n\rangle\,,
\ee
exactly as in \cite{Zam}. This leads to the evolution equation for the eigenvalues $E_n$ of $H$ \cite{Zam}

\be
R^{-1}\p_tE^{(t)}_n(R)=-\langle n|T^{(t)}_{00}|n\rangle\langle n|T^{(t)}_{11}|n\rangle+\langle n|T^{(t)}_{01}|n\rangle\langle n|T^{(t)}_{10}|n\rangle\,.
\ee

Before proceeding, it is worth doing some dimensional analysis. In a non-Lorentz invariant theory, we should treat 
time $x_0$ and space $x_1$ as having separate dimensions $[x_0]$, $[x_1]$. With that convention
\be
[T_{00}]=[x_0]^{-1}[x_1]^{-1}\,,\quad [T_{01}]=[x_0]^{-2}\,,\quad [T_{10}]=[x_1]^{-2}\,,\quad[T_{11}]=[x_0]^{-1}[x_1]^{-1}\,,
\ee
and $[t]=[t']=[x_0][x_1]$. 

As usual we have $\langle n|T_{00}|n\rangle=E_n/R$, 
$\langle n|T_{11}|n\rangle$ is the pressure $\p_RE_n(R)$, and
$\langle n|T_{10}|n\rangle=iP_n/R$, where $P_n=k_n/R$ with $k_n\in2\pi\mathbb Z$.
For a Lorentz invariant theory, the energy flux 
$\langle n|T_{01}|n\rangle$ also equals $iP_n/R$, but  in general this will have a more complicated dependence on both $P_n$ and the particular state $n$. 

For a state with a dispersion relation $E=\omega_n(P)$ in infinite volume, it is reasonable to assume that the energy flux is the product of the energy density with the group velocity:
\be\label{1.12}
\langle n|T_{01}|n\rangle=i(\omega_n(P_n)/R)\omega_n'(P_n)\,.
\ee
As we show below, this is in fact the only form for which the spectrum for large enough $R$ is independent of $t$, as expected, but in principle the matrix element could also depend on $t$ for finite values of $(R/E_nt)$. We assume that this is not the case,
in order to make progress, although for states in which the expectation value of the energy current vanishes (\em e.g. \em states with $P_n=0$) this does not play a role.

For a Lorentz invariant theory, with $\omega(P)=\sqrt{M^2+P^2}$, (\ref{1.12}) gives $iP_n/R$ as expected. However for a non-relativistic Galilean invariant theory
\be
\omega(P)=\omega(0)+P^2/2M\,,
\ee
where $\omega(0)$ is the energy in the rest frame (possibly including a chemical potential) and $M$ is the inertial mass. Thus, according to this assumption, 
\be
\langle n|T_{01}|n\rangle=i(\omega_n(0)+P_n^2/2M_n)(P_n/M_nR)\,,
\ee
and thus for a gapless state we get $iP_n^3/2M_n^2R\propto R^{-4}$.

However this scaling behavior in the gapless case is more general and is independent of the detailed form of the energy current. This is because at such a critical point with non-trivial dynamic scaling (often referred to as Lifshitz, even though this actually refers to spatially anisotropic scaling), we have $[x_0]\sim [x_1]^z$, where $z$ is the dynamic exponent, and therefore by dimensional analysis we have
\be
\langle n|T_{01}|n\rangle\propto iR^{-2z}\,,
\ee
where the constant of proportionality in general depends on the state. (In principle, it could also depend on $t$ but again we assume this not to be the case.)

With this input, the evolution equation for a general state is
\be\label{1.17}
\p_tE_n(R)=-E_n(R)\p_RE_n(R)-(P_n/R)\omega_n(P_n)\omega_n'(P_n)
\ee
where for a gapless  state the last term is $\propto R^{-2z-1}$.

For $R^{z+1}\gg t$ we expect that $E_n(R)\sim \omega_n(k_n/R)$, independent of $t$, and indeed we see from (\ref{1.17}) that this is the case, further justifying the ansatz (\ref{1.12}).

In general we expect that
\be
E_n^{(t)}(R)=f^{(t)}\,R+o(R)\,,
\ee
and we then find, as for the relativistic case \cite{Zam}, 
\be
f^{(t)}=\frac{f^{(0)}}{1+f^{(0)}\,t}\,,
\ee
and that the $o(R)$ term also satisfies (\ref{1.17}).

\begin{figure}
\centering
\includegraphics[width=0.5\textwidth]{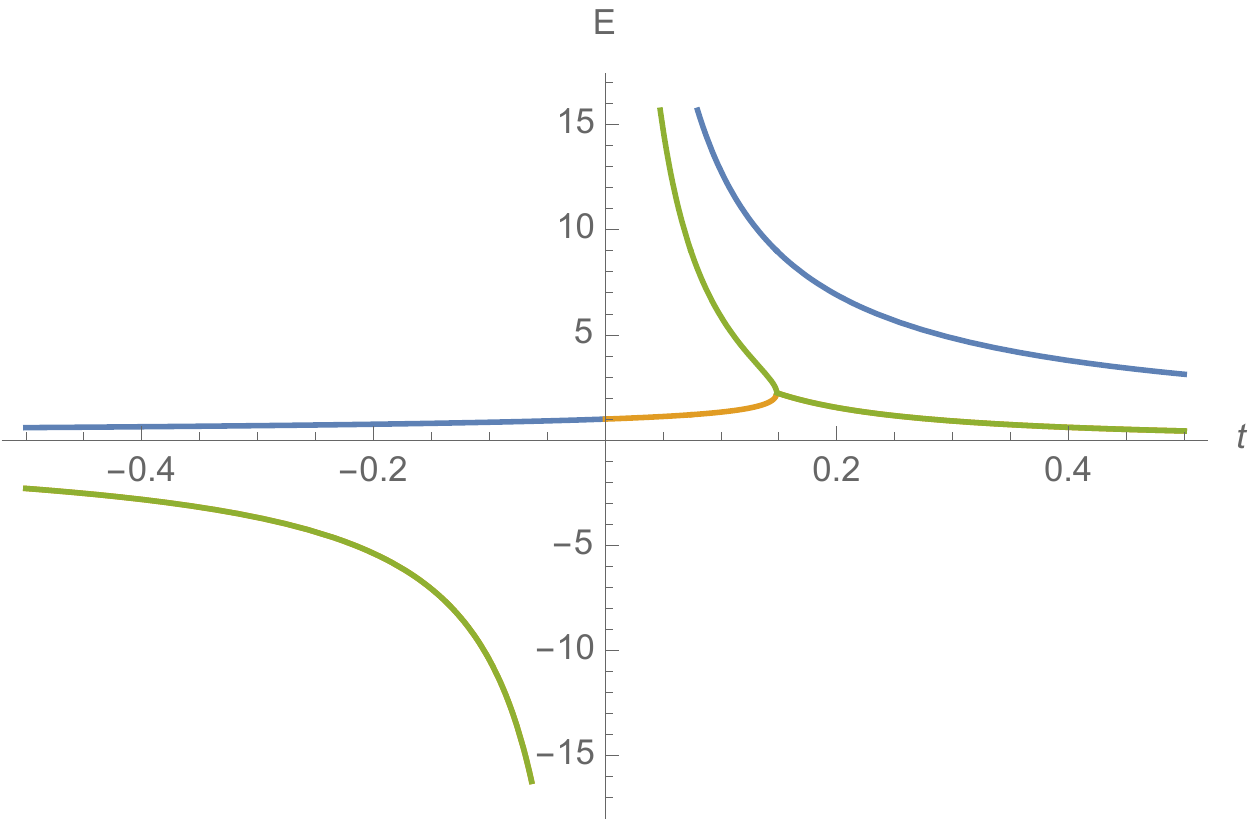}
\caption{Real parts of solutions of  (\ref{0.20}) for $z=2$ and $R=C=1$. The physical solution has a square root singularity at $t\approx 0.15$. However for $t<0$ there are complex conjugate solutions with lower energy. The solutions for $C=-1$
may be visualized by reflecting the figure in both axes. \label{fig1}}
\end{figure}

When $P_n=0$ this is the inviscid Burgers equation as in the relativistic case, with the implicit solution
\be
E_n^{(t)}(R)=E_n^{(0)}\big(R-tE_n^{(t)}(R)\big)\,.
\ee
For a gapless state at $P=0$ we have $E_n^{(0)}(R)=C_nR^{-z}$, which gives the implicit equation
\be\label{0.20}
E_n^{(t)}(R)\big(R-tE_n^{(t)}(R)\big)^z=C_n\,.
\ee 
While for integer $z$ this equation has $z$ (possibly complex) solutions, and more for fractional values, it can be shown that the physical solution becomes singular at some finite $t>0$ if $C_n>0$ and at some finite $t<0$ if $C_n<0$. In both cases,
this occurs by collision with another single zero, so the singularity is of square root type, as in the Lorentz-invariant case
with $z=1$. However, it is interesting to note that while this solution is non-singular for $t<0$, there are
then in general complex solutions whose real part actually may correspond to a lower free energy. 
This is illustrated for the case $z=2$ in Fig.~\ref{fig1}. A similar effect happens if $C_n<0$ with the reverse sign of $t$. 

In general (\ref{1.17}) may be solved by the method of characteristics:  consider
$u(t)\equiv E^{(t)}(r(t))$ so that
\be
du/dt=\p_tE+(dr/dt)\p_RE\,.
\ee
If we then choose $dr/dt=u$, then $du/dt=-V'(r(t)$, where $-V'(R)$ is the last term on the right hand side of (\ref{1.17}).
This of course simply describes a Newtonian particle of velocity $u$ in a potential $V(r)$. By quadrature
\be\label{2.22}
\ffrac12u^2+V(r)=\ffrac12(dr/dt)^2+V(r(t))={\rm const.}
\ee
 We should solve for $r(0)$ given $r(t)=R$, then
\be
E^{(t)}(R)=\left(E^{(0)}(r(0))^2+2V(r(0))-2V(R)\right)^{1/2}\,.
\ee 
If $P=0$, so $V=0$, the solution is simple: $r(t)=R=r(0)+tE^{(0)}(r(0))$, so if $E^{(0)}>0$, $r(0)\to0$ for some $t<0$, and \em vice versa\em. But when $V>0$ the particle is repelled from the origin and will reach it only if it starts with sufficient kinetic energy.  As an example, consider the gapless case with $V(R)=D/R^{2z}$ and $E^{(0)}(R)=C/R^z$. 
For $P=0$ and $C>0$ the particle reaches the origin at some $t<0$, corresponding to the square root singularity in $E^{(t)}(R)$ already discussed. However for $D>0$ it will reach the origin only if $C>(2D)^{1/2}$. Thus the higher momentum modes do not become singular. This may be seen by explicit solution of (\ref{2.22}).

\section{Torus partition function}
We may also treat partition functions using the path integral methods of \cite{Car2}. In fact the discussion is almost identical so we summarize it only. The quadratic deformation may be decoupled by a gaussian transformation
\be
e^{2\delta t\int_{\cal D}\e_{ik}\e_{jl}T^{ij}T^{kl}d^2x}\propto \int[dh]e^{-(1/8\delta t)\int\int_{\cal D}\e^{ik}\e^{jl}h_{ij}h_{kl}d^2x
+\int_{\cal D}h_{ij}T^{ij}d^2x}\,,
\ee
where the integration is over a $2\times 2$ matrix field $h_{ij}$ (not necessarily symmetric). The gaussian integral is dominated by its saddle-point $h=h^*$, which  is explicitly given by 
\be\label{3.2}
T_{00}=(1/4\delta t)h^*_{11}\,,\quad T_{11}=(1/4\delta t)h^*_{01}\,,\quad
T_{01}=-(1/4\delta t)h^*_{10}\,,\quad T_{10}=-(1/4\delta t)h^*_{01}\,.
\ee
The conservation equations $\p_jT_{ij}=0$ then imply
\be\label{3.3}
h^*_{11,0} = h^*_{10,1}\,,\quad
h^*_{00,1} = h^*_{01,0}\,,
\ee
so that
\begin{eqnarray}
&&h^*_{00} = 2\a_{0,0}\,,\quad h^*_{01} = 2\a_{0,1}\\
 &&h^*_{11} = 2\a_{1,1}\,,\quad h^*_{10} = 2\a_{1,0}\,,
 \end{eqnarray}
 where $\a_0$, $\a_1$ are arbitrary differentiable functions. Thus
 \be
 h^*_{ij}=2\a_{i,j}\,.
 \ee
The action at the saddle is then 
 \be
(1/2\delta t)\int\int\e^{ik}\e^{jl}\a_{i,j}\a_{k,l}d^2x
-\int\a_{i,j}T^{ij}d^2x 
=\int\p_j[(1/2\delta t)\e^{ik}\e^{jl}\a_i\a_{k,l}-\a_iT^{ij}]d^2x\,,
\ee
which is a total derivative. For a torus, since only $h$ needs to be single-valued and not necessarily $\a$, as explained in more detail in \cite{Car2} the only contribution comes from large diffeomorphisms with 
$2\a_{i,j}=h_{ij}=$ constant. This leads to the evolution equation for the partition function
\be\label{3.15a}
\p_t Z=2A^{-1} \e_{ik}\e_{jl}\frac{\p^2Z}{\p h_{ij}\p h_{kl}}\,,
\ee
where $A$ is the area of the torus. This may be related to the response of the partition function to changing its moduli, although, as discussed in \cite{Car2,Dub2} the argument is slightly subtle. If the torus is thought of as a parallelogram with corners at $(0,L,L',L+L')$ where $L$ and $L'$ are 2-vectors with
$A=L\wedge L'>0$, then (\ref{3.15a}) becomes 
\be\label{3.27b}
\p_tZ
=(\p_L\wedge\p_{L'})Z
-(1/A)(L\cdot\p_L+L'\cdot\p_{L'})Z\,.
\ee
This equation is identical to that found for the Lorentz invariant case, because the deformation has the same form. However, although the differential operator on the right hand side is invariant under simultaneous rotations of $L$ and $L'$, and also modular invariant, since the initial condition $Z^{(0)}$ in general breaks both of these so does the solution. Note that both sides of (\ref{3.27b}) have dimensions $[x_0]^{-1}[x_1]^{-1}$. 

\begin{figure}
\centering
\includegraphics[width=0.5\textwidth]{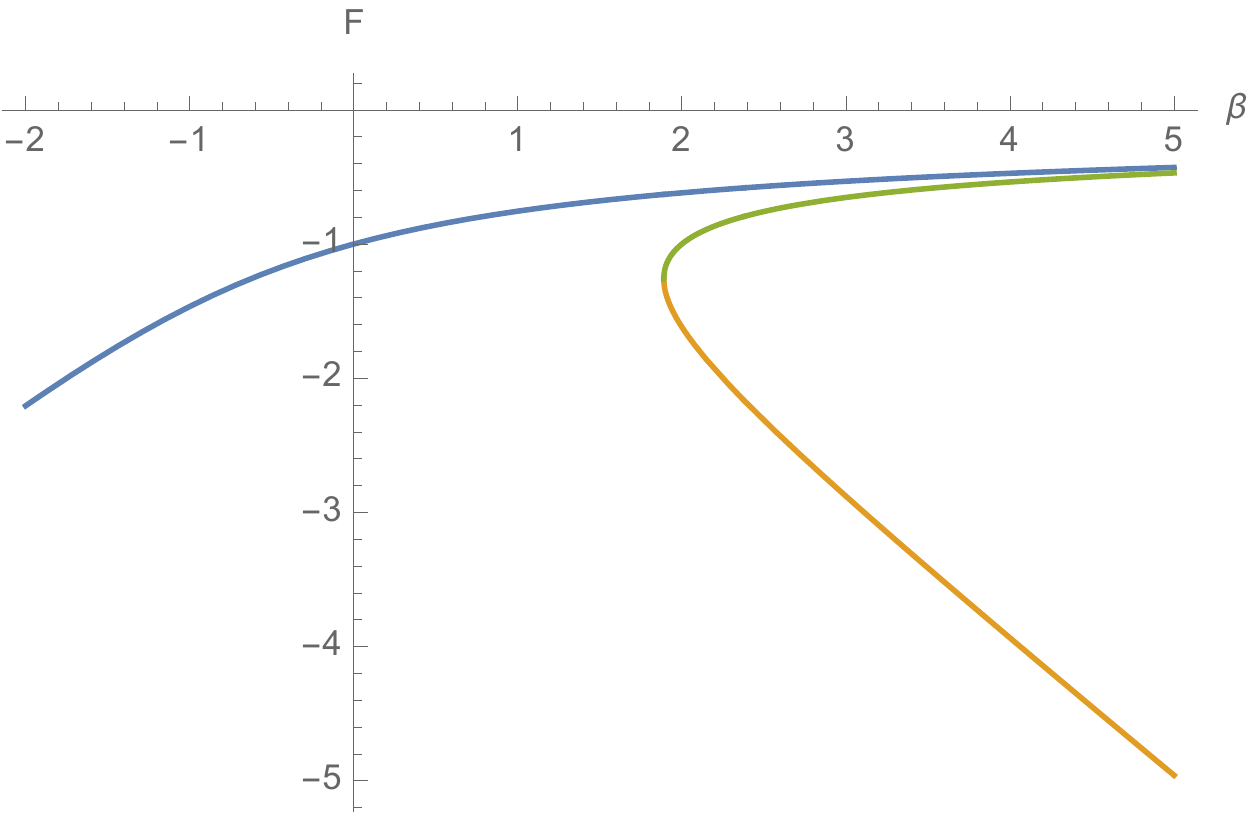}
\caption{Free energy $F$ per unit length \em v. \em inverse temperature $\beta$ for a deformed gapless system with $z=2$. The lower curve corresponds to fixed $t<0$ and exhibits the typical square root singularity of the Hagedorn transition. The upper curve is for $t>0$: the free energy and internal energy are analytic at $\beta=0$, so a continuation to negative temperature makes sense. \label{fig2}}
\end{figure}

If we now apply this symmetry to the case of rectangular torus with $L=(0,L)$ and $L'=(\beta, 0)$ we come to the conclusion that the eigenvalues of $\int T_{11}dx_0$ also satisfy (\ref{1.17}). In particular
in the limit when $L\gg\beta^{1/z}$, 
\be
Z^{(t)}\sim e^{-Lf^{(t)}(\beta)}\quad\mbox{where}\quad\p_tf^{(t)}=-f^{(t)}\p_\beta f^{(t)}\,.
\ee
In a gapless theory where $f^{(0)}(\beta)\sim-c\beta^{-1/z}$, with $c>0$ for convexity, it follows that
if $t<0$ there is a Hagedorn-type transition at some finite temperature $\beta^{-1}$ at which the free energy has a square root singularity, just as for the Lorentz-invariant case. For $t>0$, the free energy is perfectly regular as a function of $\beta$ and in fact has finite slope at $\beta=0$ corresponding to the energy density saturating at infinite temperature. 
There is then a smooth continuation to negative temperature $\beta<0$. This is illustrated in Fig.~\ref{fig2}.

Since the equation (\ref{3.27b}) is linear, it is satisfied term by  term when $Z$ is expressed as a trace over eigenstates of $H$ and $P$. This allows us to recover (\ref{1.17}). It is sufficient to consider the case when $|L_0|\ll L_1$ and $|L_1'|\ll L_0'$.
Then a typical term in the expansion has the form 
\be\label{3.11}
z=e^{-(L'_0-L_0L_1'/L_1)E^{(t)}(L_1)+\langle T_{10}\rangle L_1L_1'
+\langle T_{01}\rangle L_0L_0'}
\ee
where $\langle T_{10}\rangle=P/L_1=k/L_1^2$. In this limit we have
\be
\p_tz\sim(\p_{L_1}\p_{L'_0}-\p_{L_0}\p_{L_1'})z-(L_1L_2')^{-1}(L_1\p_{L_1}+L'_0\p_{L'_0})z\,,
\ee
and, inserting (\ref{3.11}), we find (\ref{1.17}) after a little algebra.

\subsection{Open boundaries}
Now consider a theory on the finite interval $0<x_1<L$ with open boundary conditions to be specified below. At finite temperature the partition function is given by the euclidean path integral on a finite $L\times L'$ cylinder withe periodic boundary conditions in imaginary time $x_0$. This was discussed for the relativistic case in \cite{Cav,Car2}. In this case the saddle point $h^*_{ij}(x_0,x_1)$ is independent of $x_0$ but not necessarily $x_1$. However from ({\ref{3.3}) this implies that the components $h_{00}^*$  and $h_{10}^*$ are also independent of $x_1$, and therefore constant, as are $T_{11}$ and $T_{01}$, from (\ref{3.2}). If we now require that the energy current $T_{01}$ vanishes at the boundary, then in fact
$h_{10}^*\propto T_{01}=0$ everywhere. The saddle point action is thus
$\propto h_{00}^*\int h_{11}^*dx_1$ and thus depends only on the uniform mode of $h_{11}^*$. 

We therefore come to the conclusion, as for the relativistic case, that the integration is localized on uniform values of $h_{11}$ and $h_{22}$, with $h_{10}=0$. This leads to the evolution equation \cite{Car2}
\be
\p_tZ=(4/LL')\frac{\p^2Z}{\p h_{00}\p h_{11}}=(\p_{L'}-(1/L'))\p_LZ\,,
\ee
In \cite{Car2} it was shown that if we then write $Z$ as a sum of terms of the form 
$e^{-L'E^{(t)}(L)}$ then $E^{(t)}(L)$ once again satisfies the inviscid Burgers equation, (\ref{1.17}) with $P=0$.
Note that in this case the precise form of the energy current is immaterial, and therefore this result is independent of the ansatz (\ref{1.12}).

\section{\boldmath $S$-matrix}

As for the relativistic case, the deformation implies a CDD factor dressing of the $S$-matrix. To see this for the 2-body $S$-matrix, we may generalize the argument in \cite{Smi}. 

For large enough $R$ compared with all other length scales, we assume that there are single-particle states with energies $E(p)=\omega(p)$, where $p=k/R$, $k\in2\pi\mathbb Z$, and also 2-particle states with zero total momentum and energies $E_2^{(t)}(p)\sim 2\omega(p^{(t)})$, where now $2p^{(t)}$ is the relative momentum,  quantized according to 
\be
p^{(t)}R+\Delta^{(t)}(p^{(t)})=k\in2\pi\mathbb Z\,,
\ee
where $\Delta^{(t)}(p)$ is the 2-body phase shift. Substituting this into (\ref{1.17}) leads, after a little algebra, to
\be
\p_t\Delta^{(t)}(p)|_p=-2\omega(p)\,p
\ee
so that
\be
\Delta^{(t)}(p)=\Delta^{(0)}(p)-2t\omega(p)p\,.
\ee
This corresponds to a dressing of the 2-particle $S$-matrix by a CDD factor $e^{-2it\omega(p)p}$, thus generalizing the relativistic result by the replacement $p_0\to\omega(p)$. For example, for non-relativistic potential scattering,
\be\label{4.4}
\Delta^{(t)}(p)=\Delta^{(0)}(p)-tp^3/M\,.
\ee
where $M$ is the inertial mass. 

\section{Other deformations}
Although we have focussed on the $\det T$ deformation, in fact much of the analysis extends straightforwardly to any pair of conserved currents $(J,J')$ in 1+1 dimensions, not necessarily assuming Lorentz invariance. 

For if $\p_jJ_j=\p_kJ'_k=0$, we can write, locally
\be
J_j=\e_{jm}\p_m\chi,\quad J'_k=\e_{kn}\p_n\chi'\,,
\ee
so that an infinitesimal  deformation proportional to
\be
\e_{jk}J_j(x)J'_k(x)=\e_{jk}\e_{jm}\e_{kn}(\p_m\chi)(\p_n\chi')=\e_{mn}(\p_m\chi)(\p_n\chi')=\p_m(\e_{mn}\chi\p_n\chi')=\p_n(\e_{mn}\chi'\p_m\chi)\,,
\ee
is therefore a total derivative. On a torus, because $\chi$ and $\chi'$ are not necessarily single-valued, it will integrate up to
$$
\int \langle\e_{jk}J_j(x)J'_k(x)\rangle d^2x\propto (1/A)\int\int\langle\e_{jk}J_j(x)J'_k(x')\rangle d^2xd^2x'
\propto \langle\e_{ab}Q_aQ'_b\rangle
$$
where $Q_a$, $Q'_b$ are charges flowing around each cycle $(a,b)$ (there are also some geometrical factors).

Equivalently, following Zamolodchikov, we can write 
$$
\p_{y_m}\e_{jk}J_j(x)J'_k(x+y)=(\p_{y_j}\e_{mk}+\p_{y_k}\e_{jm})J_j(x)J'_k(x+y)
$$
$$
=\e_{mk}J_j(x)\p_{x_j}J'_k(x+y)
=\p_{x_j}(\e_{mk}J_j(x)J'_k(x+y))
$$
so that, in an eigenstate of $H$ and $P$,
$$
\langle n|\e_{jk}J_jJ'_k|n\rangle=\sum_{n'}\e_{jk}\langle n|J_j|n'\rangle\langle n'|J'_k|n\rangle
$$
where the sum is over states $n'$ degenerate with $n$. If $Q_0$ and $Q_0'$ commute with $H$ and $P$ we can simultaneously diagonalize them and assume that $n'=n$.

For this to iterate then $Q_0$ and $Q_0'$ must also commute with the deformation. If they generate a symmetry it is sufficient that 
the deformation be invariant under this symmetry. Besides taking $J_j=T_{0j}$, $J_j'=T_{1j}$, corresponding to the the $\det T$ deformation, we could consider $J_j$ as generating an internal U$(1)$ symmetry and $J_j'=T_{1j}$ or $T_{0j}$. This violates parity or time reversal, but satisfies the conditions above. For the relativistic case it has been considered in \cite{Gui,Aha3}.
Similarly in \cite{Smi} examples were considered when $J$ or $J'$ represent higher spin currents.

\section{Summary}
In summary, we have shown that the solvability of the $\det T$ (``$T\overline T$'') and similar deformations of two-dimensional theories extends straightforwardly to non-Lorentz invariant theories. This includes a large of number of interesting examples of lagrangian field theories which possess a local stress-energy tensor, for example Lifshitz-type theories, non-relativistic fluids, and classical stochastic field theories such as relaxational dynamics, reaction-diffusion systems, the KPZ equation, and directed percolation, to name just a few \cite{Tau}. We showed that in all these cases, the finite-size spectrum obeys an evolution equation similar to that in the relativistic case,
in fact identical for states in which the mean energy current vanishes. For other states it was necessary to make an ansatz for the form of the current in order to obtain explicit results. However this equation appears in general to have non-perturbative solutions whose significance is at present unclear.  In general all these deformed theories for $t<0$ have a Hagedorn-type density of states. For $t>0$ the energy density saturates at a finite value at infinite temperature, with another branch corresponding to negative temperature. 

The arguments of this paper suggest that the deformation  of \em any \em translationally invariant local hamiltonian by a term of the form
\be
T_{00}\cdot T_{11}\sim\mbox{energy density}\times\mbox{pressure}
\ee
(for zero energy flux)
should in fact possess many of the characteristics of the $T\overline T$ deformation which have been discussed in the literature in the context of relativistic theories. It should therefore be a very general feature of many physical systems in 1+1 dimensions.

\acknowledgments

This work was supported in part through funds from the Simons Foundation.

\appendix\section{Non-relativistic ideal gas}
An amusing and instructive example is afforded by taking ${\cal T}^{(0)}$ to be a non-relativistic ideal gas. The states may be labelled by the momenta $\{p_a\}$ of the particles. In the non-interacting theory we then have
\begin{eqnarray}
&&T_{00}(x)=\sum_a(p_a^2/2M)\delta(x_a-x)\,,\quad T_{10}=i\sum_ap_a\delta(x_a-x)\nonumber\\
&&T_{01}(x)=i\sum_a(p_a^2/2M)(p_a/M)\delta(x_a-x)\,,\quad T_{11}=-\sum_ap_a(p_a/M)\delta(x_a-x)\,,
\end{eqnarray}
so that
\be
\det T=-(1/2M^2)\sum_{a,b}\big(p_a^2p_b^2-p_ap_b^3\big)\delta(x_a-x)\delta(x_b-x)
\ee
Note that the singular self-interaction term with $a=b$ cancels between the two terms, as expected, leaving a 2-body interaction 
\be
-(\delta t/2M^2)\sum_{a\not=b}\big(p_a^2p_b^2-p_ap_b^3\big)\delta(x_a-x_b)
\ee
\be
=(\delta t/4M^2)\sum_{a\not=b}p_ap_b(p_a-p_b)^2\delta(x_a-x_b)\,.
\ee
which may perhaps be described as soft-core scattering. It can be checked that in perturbation theory is gives rise to a phase shift of the form (\ref{4.4}).
At the next stage of the iteration 3-body interactions are generated. It is simpler to find the full solution in second quantization, assuming a euclidean Lagrangian density of the form
\be
{\cal L}=\frac12(\phi^*\p_t\phi-\phi\p_t\phi^*)+F(X)\,,
\ee
where $X=\p_x\phi^*\p_x\phi$.
From Noether's theorem and the equations of motion
\begin{eqnarray}
T_{00}&=&\frac12(\p_t\phi)\phi^*-\frac12(\p_t\phi^*)\phi-L=-F\,,\\
T_{10}&=&(1/2)[(\p_x\phi)\phi^*-(\p_x\phi^*)\phi]\,,\\
T_{01}&=&(\p_t\phi)(\p_x\phi^*)F'+(\p_t\phi^*)(\p_x\phi)F'\,,\\
T_{11}&=&2(\p_x\phi)(\p_x\phi^*)F'-F-\frac12(\phi^*\p_t\phi-\phi\p_t\phi^*)=3(\p_x\phi)(\p_x\phi^*)F'-F\,,
\end{eqnarray}
where the last expression is up to a total derivative.
Note that if $F(X)=X$ then $T_{11}+2T_{00}=0$ as expected for a scale invariant theory with $z=2$. 

Working for simplicity in the $P\propto T_{10}=0$ sector, the deformation is $\propto T_{00}T_{11}$ so we may write the evolution equation for the Lagrangian 
\be
\p_tF=F(3X\p_XF-F)\,.
\ee
This may be solved directly by the method of characteristics, or transformed into the inviscid Burgers equation by substituting $F=X^{1/3}G$, $Y=X^{-1/3}$, with implicit solution. 
\be
F(t,X)=\frac{X}{(1-tF(t,X))^2}\,.
\ee
This gives a kind of non-relativistic version of the Nambu-Goto action. At $P\not=0$ the solution is more complicated.

\end{document}